\documentclass[aip,
reprint,%
]{revtex4-1}

\usepackage{graphicx}
\usepackage{dcolumn}
\usepackage{bm}
\begin{document}

\title{High $Q$ electromechanics with InAs nanowire quantum dots}
\author{Hari~S.~Solanki}
\email[]{hss@tifr.res.in}
\author{Shamashis Sengupta}
\email[]{shamashis@tifr.res.in}
\author{Sudipta Dubey}
\author{Vibhor Singh}
\author{Sajal Dhara}
\author{Anil Kumar}
\author{Arnab Bhattacharya}
\author{S. Ramakrishnan}
\affiliation{Department of Condensed Matter Physics and Materials Science, Tata Institute of Fundamental Research, Homi
Bhabha Road, Mumbai 400005, India}
\author{Aashish A. Clerk}
\email[]{clerk@physics.mcgill.ca}
\affiliation{Department of Physics, McGill University, Montreal, Quebec, Canada H3A 2T8}
\author{Mandar~M.~Deshmukh}
\affiliation{Department of Condensed Matter Physics and Materials Science, Tata Institute of Fundamental Research, Homi
Bhabha Road, Mumbai 400005, India}
\date{\today}
\begin{abstract}
In this report, we study electromechanical properties of a suspended InAs nanowire (NW) resonator. At low temperatures, the NW acts
as the island of a single electron transistor (SET) and we observe a strong coupling between electrons and mechanical modes at
resonance; the rate of electron tunneling is approximately 10 times the resonant frequency. Above and below the mechanical resonance, the
magnitude of Coulomb peaks is different and we observe Fano resonance in conductance due to the
interference between two contributions to potential of the SET. The quality factor ($Q$) of these devices is observed $\sim$10$^5$
at 100 mK.
\end{abstract}


\maketitle

Nano electromechanical systems (NEMS) are of interest for single charge
\cite{krommer}, single spin \cite{rugar}, and spin
torque \cite{zolfagharkhani} detection studies as well as to measure mass \cite{jensen,lassagnesensor,chiu} and forces
\cite{teufel,mamin}.
NEMS have also demonstrated continuous position measurements near the fundamental limits set by the uncertainty principle
\cite{teufel,rocheleau}. All these experiments require both a strong electromechanical coupling, i.e., the coupling between the wire's motion and its charge density, as well as low-dissipation mechanical resonances; achieving both these goals in the same system can be
challenging. We demonstrate both these features (strong electromechanical coupling and high mechanical $Q$) in an InAs NW resonator.

High $Q$ is the key to achieving high NEMS sensitivity. Researchers have measured $Q$ $\sim$10$^5$ at low temperatures using
carbon nanotubes and graphene \cite{huttelQ,eichler} as an active part of the NEMS. In comparison to carbon nanotubes and graphene,
NWs offer a unique
possibility to control dimensions, composition and doping during growth process. InAs is known for high electron
mobility \cite{dayehmobility} and strong spin-orbit interaction \cite{nadj}, which make it a promising material for high
speed electronic and spintronic devices. In these experiments, we observe that our InAs NW can achieve
a $Q$ of $\sim$10$^5$ at 100 mK temperature. At low temperature, we observe sequential single electron tunneling while
the wire resonates. The mechanical motion causes a modification of Coulomb blockade current peaks; also, the tunneling of a single electron through the wire results in a dip in resonant frequency, suggesting a strong coupling between electron transport and mechanical
degrees of freedom. In addition, we observe Fano resonances in the conductance
as a function of mechanical driving frequency; these result from an interference between the change in conductance due to mechanical motion of the wire
and a direct electrostatic gating. These resonances are in good quantitative agreement with a simple model describing this interference.

\begin{figure}
\includegraphics[width=85mm]{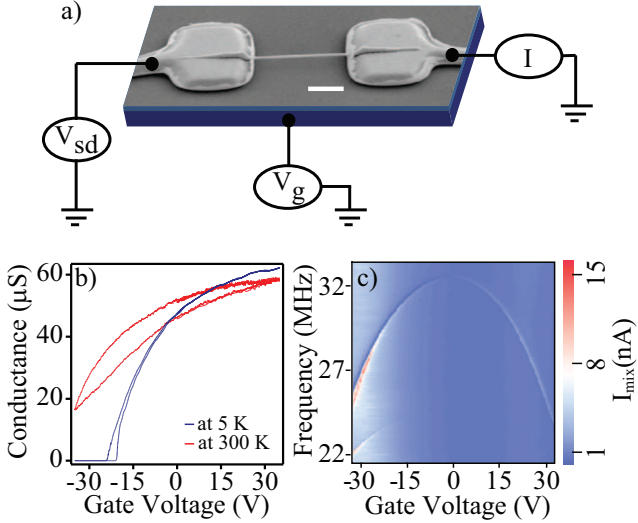}
\caption{\label{fig:figure1} (Color online) (a) Scanning electron microscope image of a typical device with circuit diagram used for
characterization (scale bar corresponds to 1 $\mu$m). (b) Tunability of NW transistor's conductance as a function of $V_g^{DC}$
at two temperatures. (c) 2D plot of mixing current shows dispersion of resonant frequency with $V_g^{DC}$ at room temperature.}
\end{figure}

Our InAs NWs are 50-150 nm in diameter and few $\mu$m long \cite{support}. We
fabricate suspended single InAs NW resonator devices in a field effect transistor (FET) geometry \cite{solanki}.
Fig. \ref{fig:figure1}(a) shows a typical device, where a NW of length $\sim$3
$\mu$m and diameter $\sim$80 nm is suspended $\sim$ 200 nm above the substrate. Fig. \ref{fig:figure1}(b) shows
conductance variation with applied DC gate voltage ($V_g^{DC}$) displaying FET behavior
at room temperature and low temperature. Room temperature
measurements of mechanical resonance are done using a heterodyne mixing technique
\cite{support,veranature,knobelrfmix}. Voltages of two different frequencies are applied to the source and gate. Consequently, the current through the device has different frequency components. The one at the difference frequency is called the mixing current ($I_{mix}$) and its magnitude depends upon the amplitude of mechanical resonance. ($I_{mix} = \frac{dG}{dV_g} (A\xi_f+B)$, where $G$ is the conductance, $V_g$ the gate voltage, $\xi_f$ the amplitude of oscillation at driving frequency $f$ applied to gate, and $A$ and $B$ are factors depending upon the magnitude of voltages applied to the source and drain.)
$I_{mix}$ as a function of actuating frequency shows a sharp change at electromechanical resonance. Fig.
\ref{fig:figure1}(c) shows 2D plot of measured mixing current as a function of frequency and $V_g^{DC}$ at 300 K. We have
observed non-monotonic resonant frequency (Fig. \ref{fig:figure1}(c)) and  mode mixing \cite{support} with $V_g^{DC}$
similar to an earlier report \cite{solanki}.

\begin{figure}
\includegraphics[width=85mm]{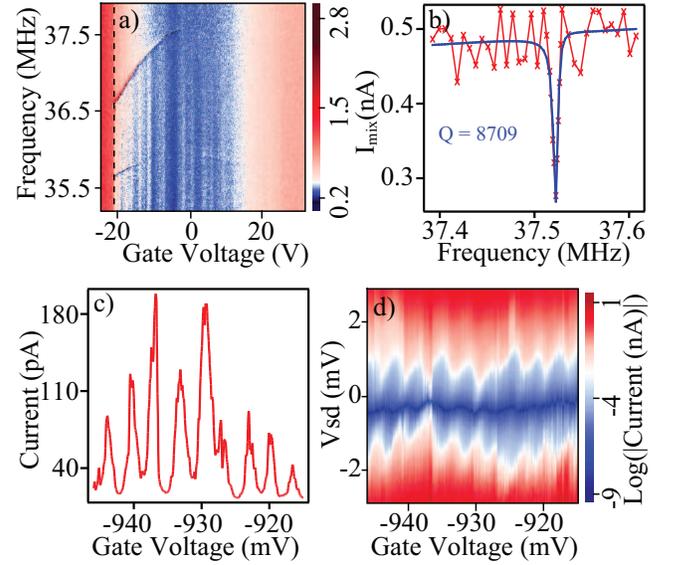}
\caption{\label{fig:figure2} (Color online) (a) Shows 2D mixing current plot as a function of driving frequency and $V_g^{DC}$
at 5K. (b) Shows $Q$ of device to be 8709 at 5K. (c) Shows Coulomb peaks in current through the NW
which corresponds to sequential single electron tunneling at 100 mK. (d) Shows 2D plot of DC current (on $log_e$ scale)
with source-drain voltage $V_{sd}$ and $V_g^{DC}$ (Coulomb diamonds). Charging energy of the system $\sim$2 meV at 100 mK.}
\end{figure}

At low temperatures, the resonant frequency of devices increases, as expected, due to variation in tension resulting from
relative contraction of NW and metal electrodes. This has been seen in other NEMS devices also
\cite{singh}. Fig. \ref{fig:figure2}(a) shows a 2D plot of mixing current as a function of frequency and $V_g^{DC}$ at 5 K.
We observe that at low temperatures, resonance features can not be detected using mixing technique after the NW FET turns off,
(towards the left of sharp vertical line in Fig. \ref{fig:figure2}(a) around $\sim-$20 V). This vertical line and the absence
of mixing current towards the left of it is a natural consequence of the mixing current being proportional to the change
in conductance with gate voltage \cite{solanki,veranature}. As the gate voltage is swept from the off-state to the on-state of the device, the conductance changes from a negligibly small value to a large value, which
leads to negligible mixing current below the on-off value of the device. From room temperature to 5K, $Q$
increases by two orders of magnitude and becomes 8709 (see Fig. \ref{fig:figure2}(b)).

Additionally, at low temperatures, our system can be viewed as a quantum dot exhibiting familiar Coulomb blockade features (Fig.
\ref{fig:figure2}(c)). In this regime, there are well defined number of electrons in the
wire and to add an additional electron, we need to provide a charging energy ($E_C=\frac{e^2}{C}$,
where $C$ is total capacitance) of the system, which is the dominant energy scale of the system at this temperature ($E_C
\gg k_BT$). At Coulomb peaks, current flows through the wire by sequential single electron tunneling \cite{steele,lassangecoupling}.
Fig. \ref{fig:figure2}(d) shows Coulomb diamonds in the current through the NW as a function of source-drain and gate voltages.
From the Coulomb diamonds (Fig. \ref{fig:figure2}(d)), charging
energy can be estimated ($E_C = e\frac{C_g}{C}\Delta V_g)$ $\sim$2 meV at 100 mK.

\begin{figure}
\includegraphics[width=85mm]{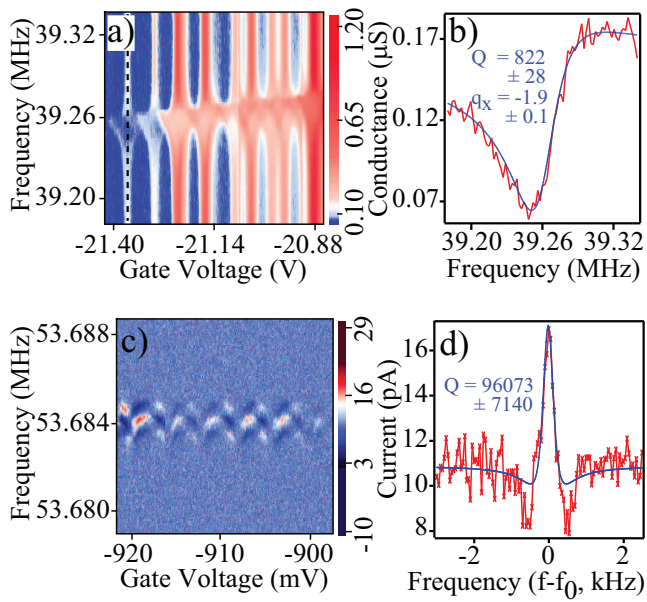}
\caption{\label{fig:figure3}(Color online) (a) Plot of the rectified signal through the device as a function of driving
frequency and $V_g^{DC}$ at 1.4 K.
(b) Line plot of conductance of the device as a function of frequency at fixed $V_g^{DC}$ (-21.36 V). Left and right side,
far from the resonant frequency, rectification signal is not same. Blue trace shows the fitting for Fano resonance. Right
at Coulomb peak $Q$ reduces to $\sim$ 800 from $\sim$ 5000.
(c) Modulated current
through the device with frequency and $V_g^{DC}$ sweep at 100 mK temperature. (d) Measured $Q$ at 100 mK comes out to be
$\sim$10$^5$ and resonant frequency, f$_0$ is 53.6838 MHz.}
\end{figure}

From Fig. \ref{fig:figure2}(a), we see that these NW devices are mechanically resonating, and also from Fig.
\ref{fig:figure2}(c) and (d), that they act as SETs. Combining these two effects allow us to study the coupling between
mechanical and charge degrees
of freedom \cite{steele,lassangecoupling}. First, we use a rectification technique \cite{steele,Resenlattthesis} to probe
Coulomb blockade physics at mechanical resonance \cite{support}. Fig. \ref{fig:figure3}(a) shows 2D plot of current through
the device at small bias voltage
($V_{sd}=100~\mu V$) at 1.4 K with $V_g^{DC}$ and frequency of RF signal. In Fig. \ref{fig:figure3}(a),
horizontal features indicate mechanical resonance
of the device and the vertical lines correspond to the Coulomb blockade resonances (Coulomb peaks). These Coulomb blockade
resonances are strongly modified when the mechanical driving frequency equals the resonant frequency.
Due to mechanical resonance, Coulomb peaks become broader irrespective of frequency and gate voltage sweep direction.

Magnitudes of Coulomb-blockade peak current is different above and below the mechanical resonance (see Fig. 3(b)) \cite{support}, which
can be understood as arising from the
interference of two terms, one is direct coupling of dot potential with applied RF actuation voltage and the other is
modification in dot potential due to mechanical motion of the dot, which leads to Fano resonance \cite{clerk,rodrigues}.
Fano resonance can be derived \cite{support} as an interference between two contributions determining
the dot potential and as a result the charge state. Briefly, the total conductance $G_{total}$, of the device will have an
additional contribution $G_{rect}$, due to the mechanical resonance besides the normal conductance $G_0$. As the wire oscillates,
its capacitive coupling with the back gate will also change at the same frequency, thereby modulating the dimensionless
gate voltage $N$, in the wire.


\begin{eqnarray}
N(t)&=C_g(t)V_g(t)/e=N_0+\delta N(t),\nonumber\\
G_{total}&=G_0+G_{rect}\simeq G_0+\frac{1}{2}\frac{d^2G}{dN^2}\overline{(\delta N(t))^2},\label{eqn.equation1}
\end{eqnarray}

where $N_0$ is its value at $V_{ac}=0$ and $\overline{(\delta N(t))^2}$ denotes time averaged quantities. $\delta N$ can be expressed
in terms of the mechanical amplitude, and we arrive at the following result for the rectification conductance

\begin{equation}
G_{rect}(\tilde\omega)=G_D\frac{|~\tilde\omega+q_x+iq_y~|^2}{\tilde\omega^2+1},\label{eqn.equation2}
\end{equation}

where $\widetilde{\omega}=\frac{\omega-\omega_0}{\gamma/2}$,
$G_D=\frac{1}{2}\frac{d^2G}{dN^2}\left(\frac{C_0}{e}\right)^2|V_{ac}|^2$,
$q_x=-\left(\frac{1}{C_0}\frac{dC}{dz}\right)\left(V_0\frac{dF}{dV_g}\right)\frac{Q}{k_{osc}}$ and
$q_y=1$. (The quantity ($q_x+iq_y$) is known as the Fano factor.) Here $\omega_0$ is the resonant frequency, $\omega$ is the frequency of applied AC signal, $\gamma$ is the damping
coefficient, $z$ is the separation between NW and substrate which depends upon the amplitude of oscillation. $k_{osc}$ is
spring constant
of the resonator ($k_{osc}=m\omega_0^2,~m$ is mass of resonator), $C$ is capacitance between the NW and the gate terminal,
$F$ is the driving force on the resonator
and $V_g$ is the total voltage applied at the gate. Fig. \ref{fig:figure3}(b) shows the conductance, right at the Coulomb peak,
as a function of frequency at fixed $V_g^{DC}$ = -21.36 V. By fitting Eqn. \ref{eqn.equation1} to Fig. \ref{fig:figure3}(b)
(blue curve) gives the real part of Fano factor ($q_x$) -1.9 while its calculated \cite{support} value is -2.7 and $q_y=1$.
While other groups have seen similar interference effects in electromechanical systems, this is, to our knowledge, the first time
that a quantitative analysis has been done. From the
fitting $Q$ of the device right at the Coulomb peaks comes
$\sim$800 while away from Coulomb peaks its value is $\sim$5000, implying that due to single electron tunneling, damping in
the system increases.

In another set of complementary measurements at lower temperatures, we measured the dependence of the mechanical resonance
frequency with $V_g^{DC}$ using
frequency modulation (FM) mixing technique \cite{gouttenoire,eichler} and found a dip in
mechanical resonant frequency at the Coulomb peak as seen in the rectification signal. The extent of coupling can also be
clearly seen at mK temperatures where a FM technique is used. Fig. \ref{fig:figure3}(c), shows
the 2D plot of mixing current through the device using FM
technique which shows dip in mechanical resonant frequency (this can also be seen by subtracting the background from the data
shown in Fig. \ref{fig:figure3}(a) \cite{support}). Similar dispersion has been seen in carbon nanotubes
\cite{steele,lassangecoupling} also. It can be understood in terms of occupation number (charge) fluctuation at Coulomb peaks
which lead to change in potential of the dot and hence softening in spring constant
occurs which is seen as a dip in resonant frequency \cite{steele,lassangecoupling}. In our device, the tunnel rate at Coulomb
peak ($I=e\Gamma$, where $\Gamma$ is tunnel rate) is
$\sim$10 times of mechanical vibration while in the previous measurements on CNT \cite{steele,lassangecoupling}, this ratio
is $\sim$300. Thus we are in different coupling regime compared to carbon nanotube resonators.
Fig. \ref{fig:figure3}(d) shows a fitting function for mixing current (using FM) through the device which shows a $Q$ of $\sim$10$^5$ at 100 mK, comparable to the highest $Q$ seen in synthesized nanostructures\cite{gouttenoire}.

In summary, we have studied InAs NW resonators FET. At low temperatures, when charging energy
dominates over thermal energy, NW acts as the island of a single SET. In this regime,
Coulomb peaks are strongly affected due to mechanical
resonance, at the same time, due to sequential electron tunneling, mechanical resonant frequency is shifted down at Coulomb peaks.
Fano resonance is observed in conductance through the NW as a function of actuation frequency due to interference in two
contributing terms in conductance. All these measurements suggest a strong coupling between electron charge and mechanical motion
of the NW. The $Q$ of these devices at low temperatures is observed to be $\sim$10$^5$. Along with strong spin
orbit interaction and a large $g$ factor of InAs, these high $Q$ and strong electro-mechanical coupling devices may allow
study of coupling mechanism between spin of electron and mechanical degree of freedom.

We acknowledge Government of India and AOARD--104141 for support.


%

\end{document}